\documentclass[useAMS,usenatbib,fleqn,a4paper]{mn2e}
\usepackage{times}
\usepackage{amsmath}
\usepackage{amssymb}
\usepackage{amsbsy}
\usepackage{bmpsize}
\usepackage{graphicx}
\usepackage{subfigure}
\usepackage{paralist}
\usepackage{mathrsfs}
\usepackage{booktabs}
\usepackage{tabularx}
\usepackage{color}

\usepackage{color}
\definecolor{darkred}{rgb}{0.55, 0.0, 0.0}
\definecolor{darkblue}{rgb}{0.0, 0.0, 0.55}
\usepackage[colorlinks=true,linkcolor=darkred,citecolor=darkblue]{hyperref}

\RequirePackage{snapshot}
\graphicspath{{}{./figs_used/}}

\def\df{{\sc df}}

\def\d{{\rm d}}

\def\degrees{\,\degr}

\def\percent{\text{ per cent}}
\def\half{{\textstyle{\frac12}}}

\def\third{{\textstyle{\frac13}}}

\def\upifour{{\textstyle{\frac\upi4}}}

\newcommand{\bs}[1]{\bmath{#1}}

\def\percent{\text{ per cent}}

\title{Self-consistent triaxial models}
\author[J. L. Sanders and N. W. Evans]{Jason L. Sanders$^1$\thanks{E-mail: jls,nwe@ast.cam.ac.uk} and N. Wyn Evans$^1$\\
$^1$Institute of Astronomy, Madingley Road, Cambridge, CB3 0HA}

\pagerange{\pageref{firstpage}--\pageref{lastpage}} \pubyear{2015}

\begin{document}
\maketitle
\label{firstpage}
\begin{abstract}
We present self-consistent triaxial stellar systems that have analytic
distribution functions (\df s) expressed in terms of the actions. These
provide triaxial density profiles with cores or cusps at the centre.
They are the first self-consistent triaxial models with analytic \df s
suitable for modelling giant ellipticals and dark haloes.
Specifically, we study triaxial models that reproduce the Hernquist
profile from Williams \& Evans (2015), as well as flattened isochrones
of the form proposed by Binney (2014). We explore the kinematics and
orbital structure of these models in some detail. The models typically
become more radially anisotropic on moving outwards, have velocity
ellipsoids aligned in Cartesian coordinates in the centre and aligned
in spherical polar coordinates in the outer parts.

In projection, the ellipticity of the isophotes and the position angle
of the major axis of our models generally changes with radius.  So, a
natural application is to elliptical galaxies that exhibit isophote
twisting.  As triaxial St\"ackel models do not show isophote twists,
our \df s are the first to generate mass density distributions that do
exhibit this phenomenon, typically with a gradient of $\approx
10^\circ$/effective radius, which is comparable to the data.

Triaxiality is a natural consequence of models that are susceptible to
the radial orbit instability. We show how a family of spherical models
with anisotropy profiles that transition from isotropic at the centre
to radially anisotropic becomes unstable when the outer anisotropy is
made sufficiently radial. Models with a larger outer anisotropy can be
constructed but are found to be triaxial. We argue that the onset of
the radial orbit instability can be identified with the transition
point when adiabatic relaxation yields strongly triaxial rather than
weakly spherical endpoints.
\end{abstract}

\begin{keywords}
Galaxy, galaxies: kinematics and dynamics -- methods: analytical, numerical.
\end{keywords}

\section{Introduction}

The case for the importance of triaxiality in galactic dynamics has a
reasonably long history. Early studies by \cite{Binney1978} and
\cite{Illingworth1977} proposed that giant elliptical galaxies are
generically triaxial in shape and this hypothesis was reinforced by
the $N$-body simulations of \cite{AarsethBinney} that showed that
aspherical initial conditions relaxed to triaxial distributions. More
recently, triaxiality is believed to be a generic feature of so-called
`slow rotator' galaxies
\citep{Emsellem2007,Cappellari2007,Krajnovic2011}.

When tackling the data it becomes difficult to disentangle triaxiality
from other effects, not least because we are only ever able to observe
a projection of the distribution. However, even in projection,
triaxiality can give rise to signature effects. In particular, a
changing axis ratio of concentric ellipsoidal density contours gives
rise to a twisting of the isophotes when viewed from a general
angle. The slow rotators of the SAURON sample \citep[][those with
  $\lambda_R<0.1$]{Emsellem2007} show evidence of isophote
twisting. In addition to inspecting a single galaxy for evidence of
triaxiality, statistics on the shapes of many galaxies yield important
information. Using the assumption of random viewing angles and a
wholly axisymmetric population one can recover the distribution of intrinsic axis
ratios of a sample of galaxies. For instance,
\cite{TremblayMerritt1995} showed from a sample of $171$ bright
ellipticals that wholly oblate or prolate populations were ruled out
and a triaxial population favoured. More recently, \cite{Weijmans2014}
performed a similar procedure on the ATLAS-3D sample
\citep{Cappellari2011} and showed there were strong indications of
triaxiality in the sample of slow rotators.

The recent SAURON and ATLAS-3D projects
have complemented photometry with line-of-sight velocity distributions for samples of
nearby galaxies. This additional information helps to unravel the
internal dynamics, and hence the intrinsic shapes, of the
galaxies. \cite{Binney1985} proposed that rotation along the minor
axis of projected elliptical density contours or instead kinematic
misalignment (misalignment of the minor axis and the angular momentum
vector) is indicative of triaxiality. \cite{FranxIllingworthdeZeeuw}
inspected the statistics of minor-axis rotation for a sample of
galaxies and found $\geq35\percent$ of their sample had kinematic
misalignment. \cite{Weijmans2014} used kinematic misalignment to
separate off potential triaxial candidates in their sample.

So far our discussion of the evidence for triaxiality has been limited
to the visible component of galaxies. The shapes of dark matter haloes
remains an interesting open question. Many $N$-body simulators
\citep{JingSuto2002,BailinSteinmetz2005,Allgood2006} have demonstrated
the range of possible triaxial dark matter haloes in cosmological
simulations. However, the models relax to axisymmetry or sphericity
when baryons are included in the simulations
\citep{Udry1993,Dubinski1994,BarnesHernquist1996,Kazantzidis2004,Debattista2008}
or when a central black hole forms \citep{MerrittQuinlan1998}. The
case for triaxiality of the Milky Way's dark matter halo is motivated
by studies of the Sagittarius stream
\citep{LawMajewski2010,VeraCiroHelmi2013,DegWidrow2013}. However,
several of these results are unsettling in that the short-axis of the
dark matter halo lies in the disc plane
\citep{Debattista2013}. Unhappily, the Sagittarius stream has proved
hard to model, with no single fit able to explain the wealth of new
data.

\subsection{Dynamical models}

Although there is considerable evidence for the presence of
triaxiality in galaxies, the range of modelling tools available is
severely limited. The two greatest developments in the understanding
of triaxial stellar systems were the work of \cite{deZeeuw1985a} on
St\"ackel or separable potentials and \cite{Schwarzschild1979} on the
construction of numerical dynamical equilibria.  \cite{deZeeuw1985a}
showed that the only St\"ackel potential in which the isodensity
surfaces are concentric ellipsoids is the perfect ellipsoid and
provided key insights into the classes of orbits that are expected to
arise in generic triaxial potentials. \cite{Schwarzschild1979} showed
how a given triaxial density-potential pair could be self-consistently
generated by linear superposition of numerically integrated
orbits. \cite{Statler1987} used this method to construct numerical
dynamical models of de Zeeuw's perfect ellipsoid and demonstrated that
they generically produce kinematic misalignment when viewed in
projection. However, as shown by \cite{Franx1988}, triaxial St\"ackel
models are unusual in that they are unable to produce isophote
twisting. Schwarzschild's method has proved useful in modelling
external galaxies, but the only triaxial fit to an elliptical galaxy
remains the work of \citet{vandenBosch2008} on NGC 4365 -- though the
related problem of modelling rotating triaxial bars has seen some
progress~\citep{Zh96,Ha00,Wa13}. Triaxial models have also be
constructed using a made-to-measure approach
\citep{SyerTremaine1996,Dehnen2009}, though here the only fits to data
are for the Milky Way bar~\citep{Lo13,Po15}.

An alternative approach to the construction of dynamical equilibria is
through the use of integrals of motion. The Jeans theorem states that
an equilibrium distribution function (\df) is solely a function of the
integrals of motion. In spherical systems the classical integrals of
the energy, $E$, and angular momentum, $L$, can be used to construct
equilibrium models. For instance, Eddington inversion allows us to
construct the isotropic $f(E)$ model given any radial density profile
$\rho(r)$ and these calculations may incorporate anisotropy in a
variety of ways
\citep{Osipkov1979,Merritt1985,Ev06,BinneyTremaine}. Moving from
sphericity to axisymmetry, it has long been realised that, in addition
to the energy and a single-component of the angular momentum (e.g. the
$z$-component $L_z$), numerically integrated orbits obey a third
non-classical integral, $I_3$ \citep[e.g.,][]{Ollongren1962} and such
a dependence in the \df\ seems necessary to reproduce the dynamics of
the Milky Way and external galaxies
\citep[e.g.,][]{Binney1976,Davies1983}. In general potentials, a
global analytic third integral does not exist, but in axisymmetric
St\"ackel potentials this integral can be written down
\citep[e.g.,][]{LB62,deZeeuw1985a}. Similarly, in triaxial St\"ackel
potentials, we can find a second non-classical integral $I_2$ (a
generalisation of a component of the angular momentum) to give three
globally defined analytic integrals of motion. However, no analytic
\df s based on the three integrals have ever been constructed for
triaxial St\"ackel models, with the exception of the very idealized
case when the tube orbits are all infinitesimally thin~\citep{Hu92}.
There are some \df s known for rotating triaxial stellar systems,
though only for rather unrealistic density distributions such as
homogeneous ellipsoids or polytropes~\citep{Fr66,Va80}

\subsection{Action-based distribution functions}

As any function of the integrals of motion is also an integral of
motion, we are free to take any choice of these functions as the basis
for our \df . A natural choice is the action coordinates,
$\bs{J}$. Along with the angle variables, they form a canonical set of
coordinates. They possess the following advantages over the classical
integrals: the range of each action is independent of the other
actions, they are adiabatic invariants (crucial for the application in
this paper) and they possess a physical meaning e.g. the `radial'
action $J_r$ describes the extent of radial excursions \citep[see][for
  more information]{BinneyTremaine}. As the actions are adiabatic
invariants, we are able to propose a form for the \df\ and iteratively
converge to a self-consistent solution. However, in order to do this
we must evaluate the properties of a given $f(\bs{J})$ model so we
require routines to find $\bs{J}$ given $(\bs{x},\bs{v})$. In
spherical potentials the actions are given by
$\bs{J}=(J_r,J_\phi,J_\theta)=(J_r,L_z,L-|L_z|)$. $J_r$ can be
calculated as a quadrature for any spherical potential.  In recent
years, several algorithms have been developed to estimate the actions
in general potentials. \cite{Sanders2012a} proposed an algorithm where
an axisymmetric potential was locally fitted by a St\"ackel potential
and then the actions estimated as those in the best-fitting St\"ackel
potential. A similar routine was presented by \cite{Binney2012} who
rewrote the equations for the actions in a St\"ackel potential in such
a way that they could be applied to a general potential. The accuracy
of both of these approaches relied on the target potential being
sufficiently close to a St\"ackel potential but crucially only
locally. \cite{SandersBinney2015_Triax} built on the work of
\cite{Binney2012} by generalising the method to triaxial
potentials. In addition to these approximate methods for finding the
actions, \cite{SandersBinney2014} presented a more accurate approach
for general triaxial potentials that relied on the construction of a
generating function from time samples of an orbit integration.

Given the tools now available for the calculation of actions, it is
natural to begin constructing self-consistent action-based \df
s. \cite{Binney2014_ISO} demonstrated how the isochrone \df\ could be
flattened to create a family of axisymmetric action-based models. More
recently, both \cite{Posti2015} and \cite{WilliamsEvans2015} have
proposed families of action-based distribution functions that
reproduced the density profiles and kinematics of a range of spherical
potentials. With the tools in hand to estimate actions in triaxial
potentials, this paper makes the obvious next step of generalizing
these models to triaxiality with the aim of producing models that are
appropriate for the modelling of the triaxial structures discussed in
the previous sections. We demonstrate how triaxial self-consistent
models can be generated and inspect the properties of a few
illustrative cases.

In Section~\ref{Machinery}, we detail all the pieces of machinery
required to construct self-consistent triaxial models. Some technical
details are reserved for three appendices. In Section~\ref{Models}, we
present triaxial versions of two recent action-based distribution
functions in the literature. In Section~\ref{Applications}, we discuss
two novelties of the presented models: isophote twisting and the
ability to explore the radial stability of the models. Finally, we
discuss the importance of resonances and chaos for these models and
other uses of the models in Section~\ref{Discussion} and draw our
conclusions in Section~\ref{Conclusions}.

\section{Constructing a self-consistent triaxial
  model}\label{Machinery}

In order to find the self-consistent potential $\Phi$ for a given $f(\bs{J})$ we require a
considerable amount of machinery. Here we will outline and review the
various pieces of machinery required.

\subsection{Review of the action estimation schemes}

For the calculation of properties of a given $f(\bs{J})$, we require
algorithms for the computation of $\bs{J}$ from $(\bs{x},\bs{v})$
given some general potential. Actions only exist for integrable
potentials and chaos can play a significant role in many triaxial
potentials~\citep{Me96,Si00}. We will assume in what follows that all
orbits can be labelled by an action (or some approximate action) and
defer discussion of the presence of resonances and chaos to
section~\ref{Discussion}.

We will use two algorithms for the computation of actions. The first
is the triaxial St\"ackel fudge method introduced in
\cite{SandersBinney2015_Triax} and is appropriate for the rapid
generation of models. The second is a slower but more accurate method
using the generating function from a analytic set of angle-actions to
the target calculated from orbit integration. This second method was
detailed in \cite{SandersBinney2014} and will be useful for checking
the results of our calculations using the St\"ackel fudge.  Here, we
give an outline of the first of these methods and detail any alterations made for
the current application. The second method is detailed in Appendix~\ref{CrossCheck} where a cross-check of our calculations is presented.

\subsection{Triaxial St\"ackel fudge}

\cite{SandersBinney2015_Triax} introduced an algorithm for rapidly
estimating the actions in a general triaxial potential. It built on
the work of \cite{Binney2012} and operates by assuming the target
potential is sufficiently close to a St\"ackel potential over the
region a given orbit explores.

The Hamilton-Jacobi equations are separable in a St\"ackel potential
such that the actions can be expressed as 1D quadratures. The
canonical coordinates in which the equations separate are
$(\tau,p_\tau)$ where $\tau$ are ellipsoidal coordinates
$(\lambda,\mu,\nu)$ given by the roots of
\begin{equation}
\frac{x^2}{\tau+\alpha}+
\frac{y^2}{\tau+\beta}+
\frac{z^2}{\tau+\gamma}
=1.
\end{equation}
Here, $(x,y,z)$ are Cartesian coordinates and $\alpha,\beta,\gamma$
are parameters that describe the confocal coordinate system. We impose
$\alpha<\beta<\gamma$ such that $x$ is the long axis and $z$ the short
axis.

The most general St\"ackel potential, $\Phi_S$, can be expressed as
\begin{equation}
\Phi_S(\lambda,\mu,\nu) = \sum_{(\lambda\mu\nu)}\frac{f(\lambda)}{(\lambda-\mu)(\nu-\lambda)},
\end{equation}
where $(\lambda\mu\nu)$ denotes cyclic permutations of the three
variables. $\Phi_S$ is composed of a single function of one variable,
$f(\tau)$. The equations of motion can be written as
\begin{equation}
2(\tau+\alpha)(\tau+\beta)(\tau+\gamma)p_\tau^2=\tau^2E-\tau a+b+f(\tau)
\label{Eq::EqnOfMotion}
\end{equation}
where $E$ is the energy and $a$ and $b$ are separation
constants. $p_\tau$ is solely a function of $\tau$ such that the
actions are given by 1D quadratures
\begin{equation}
J_\tau = \frac{2}{\upi}\int_{\tau_-}^{\tau_+}\d\tau\,|p_\tau(\tau)|.
\end{equation}
$(\tau_-,\tau_+)$ are the roots of the right hand side of
equation~\eqref{Eq::EqnOfMotion}. Note that as in
\cite{SandersBinney2015_Triax} we have defined a full oscillation in
$\tau$ to be two full oscillations from $\tau_-$ to $\tau_+$ and back
again. This gives twice the true radial action for loop orbits but
means orbits continuously fill action space \citep{BinneyS}.

Given a general triaxial potential, we define the quantities
\begin{equation}
\begin{split}
\chi_\lambda(\lambda,\mu,\nu) &\equiv (\lambda-\mu)(\nu-\lambda)\Phi(\lambda,\mu,\nu),\\
\chi_\mu(\lambda,\mu,\nu) &\equiv (\mu-\nu)(\lambda-\mu)\Phi(\lambda,\mu,\nu),\\
\chi_\nu(\lambda,\mu,\nu) &\equiv (\nu-\lambda)(\mu-\nu)\Phi(\lambda,\mu,\nu).
\end{split}
\end{equation}
If $\Phi$ were a St\"ackel potential, these quantities would be given
by, for instance,
\begin{equation}
\chi_\lambda(\lambda,\mu,\nu) = f(\lambda)-\lambda\frac{f(\mu)-f(\nu)}{\mu-\nu}+\frac{\nu f(\mu)-\mu f(\nu)}{\mu-\nu}.
\end{equation}
Therefore, for a general potential, we can write
\begin{equation}
f(\tau) \approx \chi_\tau(\lambda,\mu,\nu)+C_\tau\tau+D_\tau,
\end{equation}
where $C_\tau$ and $D_\tau$ are constants provided we always evaluate
$\chi_\tau$ with two of the ellipsoidal coordinates fixed. For
instance, we always evaluate $\chi_\lambda$ at fixed $\mu$ and $\nu$.

When we substitute these expressions into equation~\eqref{Eq::EqnOfMotion}, we find
\begin{equation}
2(\tau+\alpha)(\tau+\beta)(\tau+\gamma)p_\tau^2=\tau^2 E -\tau A_\tau+B_\tau +\chi_\tau(\lambda,\mu,\nu).
\label{Eq::EqnOfMotion_JK}
\end{equation}
For each $\tau$ coordinate, there are two new integrals of motion
given by $A_\tau=a-C_\tau$ and $B_\tau=b+D_\tau$. Using a single 6D
coordinate and a choice of coordinate system gives us a single
constraint on a combination of $A_\tau$ $B_\tau$. Due to the
separability of the Hamilton-Jacobi equation, the partial derivative
of the Hamiltonian with respect to any of the ellipsoidal coordinates
is zero for a true St\"ackel potential. Setting it equal to zero for a
general potential gives a further constraint on $A\tau$ and $B_\tau$
allowing us to solve for these integrals given only a single
$(\bs{x},\bs{v})$ coordinate. We have therefore produced approximate
$p_\tau(\tau)$ equations of motion which may be integrated to estimate
the actions.

In the above algorithm, the only parameters we can control are
$\alpha$ and $\beta$ which determine the location of the foci. We set
$\gamma=-1$ without any consequences. These are chosen on an
orbit-by-orbit basis so we are free to change them every time we
require another action. We give details of the method employed to do
this in Appendix~\ref{Coords}.

\subsection{Adiabatic relaxation}

A \df\ must obey the collisionless Boltzmann equation. For some
applications such as the stellar halo, we may consider the \df\ of a
tracer population with negligible mass that lives in the potential
generated by a more dominant component. However, when treating massive
components of galaxies, it is preferable for the model to
self-consistently generate its own potential.  This poses a problem
for action-based \df s. In order to find the density
and the potential of an action-based \df , we must
know the potential in order to calculate the actions. Therefore, the
models must be constructed iteratively.

As the actions are adiabatic invariants, the form of the \df\ does not
change under slow changes of the potential. We therefore pose an
initial guess of the potential $\Phi_0$ and find the potential
$\Phi_1$ of the \df\ in $\Phi_0$. We repeat this procedure with
$\Phi_0$ replaced by $\Phi_1$ until the difference between the two
potentials is less than $\sim$ one per cent everywhere. This procedure is not
guaranteed to converge if $\Phi_0$ is very different from $\Phi_1$. To
aid with convergence \cite{Binney2014_ISO} used a linear combination of
$\Phi_0$ and $\Phi_1$ as the next trial potential. We did not adopt
that procedure here but found our models converged within $\sim 8$
iterations.

\subsection{Multipole expansion}

At each stage in the iterative procedure, we require the potential
calculated from the density. Here, we are working with general triaxial
mass distributions so we evaluate the potential using a multipole
expansion \citep{BinneyTremaine}. We discuss the details of this in
Appendix~\ref{MultiExp}.

\subsection{Tests}

The code was tested by reproducing spherical versions of the models
detailed in the next section allowing for triaxiality and comparing to
the models constructed by limiting the models to sphericity
(i.e. using only $l=0$ and $m=0$ terms in the multipole expansion and
calculating the actions in a spherical potential). Tests of the multipole expansion code are given in Appendix~\ref{MultiExp}. The action-finding
code has been tested in \cite{SandersBinney2015_Triax} and it was found that the actions were accurate to $\lesssim10\percent$ for a typical box orbit and $\lesssim5\percent$ for a typical loop orbit in a triaxial NFW potential. A comparison of the density calculation with two different action estimation approaches is presented in Appendix~\ref{CrossCheck}.


\section{Action-based models}\label{Models}

Due to the limited number of cases for which analytic
self-consistent action-based \df s can be computed the study of
action-based \df s has been fairly limited. In this section, we give
details of several recent models which will be of use in constructing
triaxial models. We consider \df s that are functions of the absolute
values of the actions. Therefore, these models do not have a streaming
velocity. This is not a necessary requirement for the models but it
simplifies their construction. In all formulae, $M$ is the mass of the
model and $\mathcal{N}$ is the normalization given by
\begin{equation}
\begin{split}
\mathcal{N} &= M^{-1}(2\upi)^3\int\mathrm{d}^3\bs{J}\,f(\bs{J})\\ &= M^{-1}(2\upi)^3\mathcal{F}\int_0^\infty\mathrm{d}J_r\int_0^\infty\mathrm{d}J_\phi\int_0^\infty\mathrm{d}J_\theta\,f(\bs{J}).
\end{split}
\end{equation}
The definition of the actions in the triaxial case (i.e. the radial
action for the loop orbits is multiplied by two) means the integral
includes loop orbits rotating in both senses so $\mathcal{F}=1$
but in the spherical case we must multiply the integral by a factor of
$\mathcal{F}=2$ to include orbits rotating both clockwise and
anticlockwise. We introduce a change of variables $s_i =
J_i/(J_0+J_i)$ to map the infinite limit to $(0,1)$ and calculate the
integral numerically using the Divonne routine in the \textsc{cuba}
package \citep{cuba} (all integrals over the \df\ are performed with this package). Additionally in the triaxial case we evaluate
the distribution function as $f(\half J_r,J_\phi,J_\theta)$ such that
we correctly map onto the spherical case.

\cite{SandersBinney2015_Triax} inspected a simple \df\ given by
\begin{equation}
f_{\rm SB}(\bs{J}) = M \mathcal{N} (J_0+\mathcal{L}(\bs{J}))^p,
\end{equation}
where
\begin{equation}
\mathcal{L}(\bs{J}) = J_r + a_\phi |J_\phi| + a_\theta |J_\theta|.
\end{equation}
Here, the coefficients $a_i$ are constants and $J_0$ is a scale
action. The models were evaluated in a fixed Navarro-Frenk-White
potential and so were not self-consistent. They have a density core and
density fall-off like $r^{-3}$ and the coefficients controlled the
kinematics. Two models were studied and it was demonstrated that the
Jeans equation was satisfied to good accuracy. These models are
simple, but lack flexibility. Here we examine two other models from
the literature which we will study further with our new machinery.

\subsection{Williams \& Evans models}

\cite{WilliamsEvans2015} proposed a family of action-based models with
double power-law density profiles of the form
\begin{equation}
\rho\propto r^a(r_0+r)^{(b-a)}
\end{equation}
and tunable anisotropy profiles. The models built on the work of
\cite{WilliamsEvans2014} who demonstrated that in scale-free spherical
potentials ($\Phi\propto r^\epsilon$), the Hamiltonian is well
approximated by a homogeneous function of degree one in the actions
i.e. $H(\bs{J})\propto(L+D_{\rm WEB}J_r)^\zeta$ where
$\zeta=2\epsilon/(\epsilon+2)$. $D_{\rm WEB}$ is a coefficient that
can be calculated by considering the limiting cases $J_r=0$ and
$L=0$. Adjusting $D_{\rm WEB}$ away from the isotropic value produces
models with constant non-zero anisotropy.

For a double power law in the density, the potential in the near and
far field is well approximated by a scale-free power law such that an
appropriate isotropic \df\ can be constructed by stitching together two
scale-free $H(\bs{J})$ as
\begin{equation}
f_{\rm WE}(\bs{J}) = \mathcal{N}M \frac{T(|\bs{J}|)\mathcal{L}(\bs{J})^{-\delta}}{\Big[J_0^2+\mathcal{L}(\bs{J})^2\Big]^{(\eta-\delta)/2}},
\label{eq:WEeq}
\end{equation}
where
\begin{equation}
\mathcal{L}(\bs{J})=L+D(|\bs{J}|)J_r.
\end{equation}
Here, $J_0$ is a scale action related approximately to the scale
radius $r_0$ of the model as $J_0=\sqrt{GMr_0}$ and
$\delta=(6-a)/(4-a)$ and $\eta=2b-3$. Both $D(|\bs{J}|)$ and
$T(|\bs{J}|)$ are transit functions of the form
\begin{equation}
\mathcal{T}(|\bs{J}|) = \frac{\mathcal{T}_0+\mathcal{T}_1|\bs{J}|/J_\mathcal{T}}{1+|\bs{J}|/J_\mathcal{T}},
\end{equation}
where $|\bs{J}|=\sqrt{J_r^2+L^2}$. $\mathcal{T}(|\bs{J}|)$ evolves
from $\mathcal{T}_0$ to $\mathcal{T}_1$ over a scale $J_\mathcal{T}$.
$T(|\bs{J}|)$ controls the weight of the near-field part of the
model relative to the far-field part. It acts to adjust the location of the
break-radius in the density profile. $D(|\bs{J}|)$ controls the
anisotropy of the model. Note that the scale-action is summed in
quadrature with $\mathcal{L}$. This was found by
\cite{WilliamsEvans2015} to produce better fits to the curvature of
the density profile around the break radius. The model presented by
\cite{WilliamsEvans2015} is very similar to that presented by
\cite{Posti2015}. However, the model of \cite{WilliamsEvans2015} is
more flexible in the range of density profiles and kinematics that can
be fitted so we adopt it here.

In the spherical case the actions $J_\phi$ and $J_\theta$ are combined
into $L=|J_\phi|+|J_\theta|$ so $J_\theta=L-|J_\phi|$. To generalize
these models beyond sphericity, we set
\begin{equation}
\mathcal{L}=D(|\bs{J}|)|J_r|+|J_\phi|+|J_\theta|/q_z.
\label{flattening}
\end{equation}
$q_z$ acts to flatten the model in the $z$-direction. Some
combinations of parameters will produce axisymmetric models, whilst
others will be triaxial, but there is no way of finding the symmetry
of the model without a full calculation.

\subsubsection{Example model}\label{WEModel}

As a demonstration of the methodology, we will construct triaxial
analogues of the \citet{Hernquist} profile using equation~\eqref{eq:WEeq}. We
refer to these as WEH models, as they use the ansatz introduced by
\cite{WilliamsEvans2015}.

A spherical Hernquist profile has the density-potential pair
\begin{equation}
\begin{split}
\rho_{\rm H}(r) &= \frac{M}{2\upi}\frac{1}{r(r_0+r)^3},\\
\Phi_{\rm H}(r) &= -\frac{GM}{r+r_0}.
\end{split}
\end{equation}
We set $G=M=r_0=1$. To construct an appropriate action-based model for
this profile, we take the parameters from \cite{WilliamsEvans2015} as
a starting point. They find an isotropic Hernquist profile is given by
$\delta=5/3$, $\eta=5$, $D_0=D_{\rm WEB}=1.814$, $D_1=1$, $T_0=0.378$,
$T_1=1$, $J_0=J_T=1$ and $J_D=0.41$. We then introduce a flattening
$q_z=0.4$. This simple alteration produces a model that is oblate in
the far-field but weakly prolate in the centre. As our action-finding
algorithm is not designed to find the actions in prolate cases, these
results are perhaps not to be completely trusted. To encourage the
model to be oblate at the centre, we increased the central radial
anisotropy by decreasing $D_0$. As noted in \cite{WilliamsEvans2015}
when altering $D_i$, $T_i$ must be adjusted accordingly to
retain the required density profile as
\begin{equation}
\begin{split}
T_0&\rightarrow T_0\Big(\frac{1+D_{\rm WEB}}{1+D_0}\Big)^{-\delta}\\
T_1&\rightarrow T_1\Big(\frac{2}{1+D_1}\Big)^{-\eta}.
\end{split}
\label{T0T1alteration}
\end{equation}

\begin{figure}
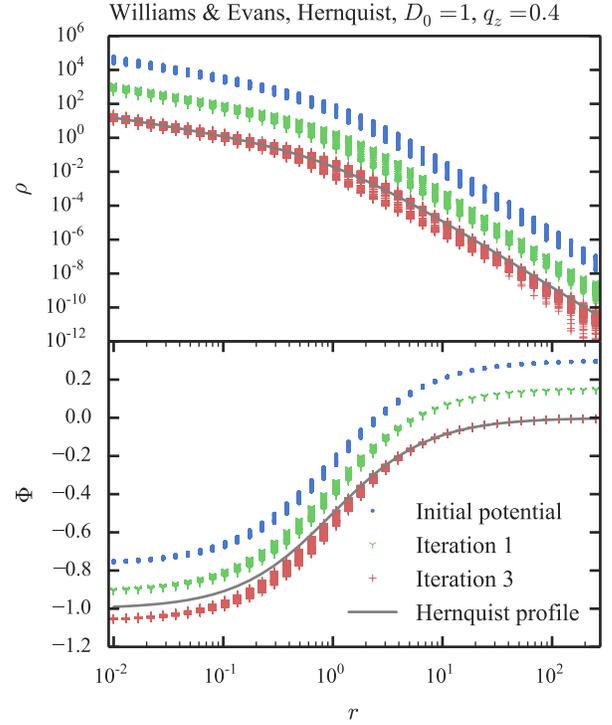

$$\includegraphics{{{figs/Fig1}}}$$
\caption{Convergence and final profiles of a triaxial WEH model: the
  top panel shows three iterations of the density calculation at
  arbitrary multiplicative offsets for visibility (the final iteration has
  no offset) and the bottom panel shows the corresponding potentials
  with arbitrary additive offsets. The blue dots show the calculation
  in the initial Hernquist potential (shown in grey), the green
  down-facing carets show the first iteration, red crosses the third
  and final iteration. Note we show the density calculated at several
  $(\phi,\theta)$ values for each spherical radius $r$ to give an
  indication of the triaxiality of the model. The positions are in
  units of $r_0$, the density in units of $M/r_0^3$ and the
  potential in units of $GM/r_0$.}
\label{ConvergenceWE}
\end{figure}

For $D_0=1$, the model is triaxial at the centre (we will discuss
under what conditions the models become triaxial later). We use as our
initial potential guess a Hernquist profile flattened in the potential
by $b/a=0.98$ and $c/a=0.95$. In Fig.~\ref{ConvergenceWE}, we show the
convergence of the $D_0=1$ model towards self-consistency.
 Note also that at each radial
point, there is a spread of points corresponding to a range of
spherical polar angles.

\begin{figure*}
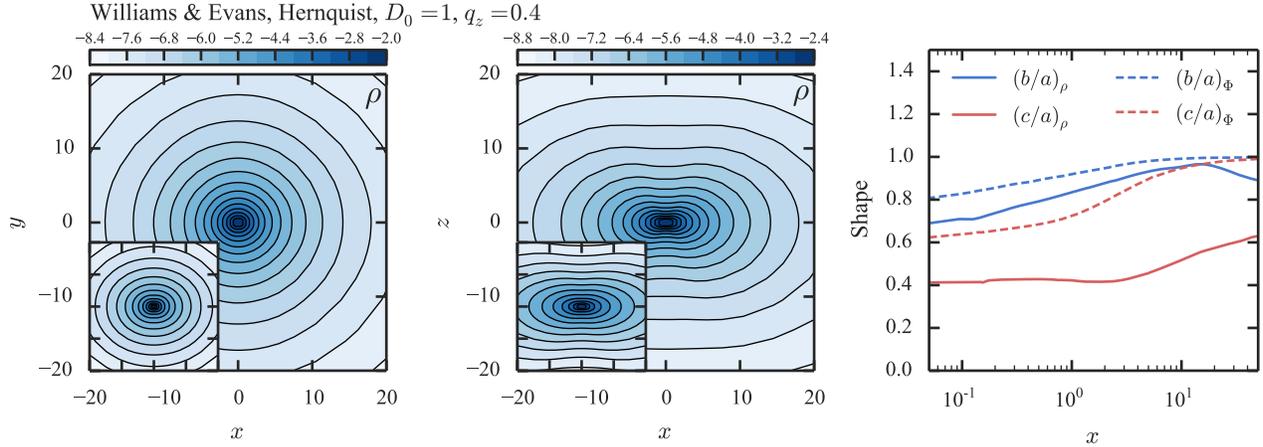

$$\includegraphics{{{figs/Fig2}}}$$
\caption{Density and shape of the triaxial WEH model: the left panel
  shows a slice of the density in the $(x,y)$ plane and the middle
  panel shows the equivalent in the $(x,z)$ plane. The blue colours
  indicate the base $10$ logarithm of the density relative to the
  central value. The insets show zoom-ins for $|x|<1,|y|<1$ and $|x|<1,|z|<1$. In
  the right-hand panel, we show the axis ratios of the fitted ellipses
  in these two planes with solid lines [$b/a$ corresponding to $(x,y)$
  and $c/a$ $(x,z)$]. The dashed lines show the equivalent for the
  potential. The positions are in units of $r_0$ and the density is in
  units of $M/r_0^3$.}
\label{DensityWE}
\end{figure*}

In Figure~\ref{DensityWE}, we plot the density contours of the model
in the $(x,y)$ plane and $(x,z)$ plane along with the axis ratios of
the best-fitting ellipses to the density and potential contours. $b/a$
is the ratio in the $(x,y)$ plane, whilst $c/a$ is the ratio in the
$(x,z)$ plane (note here we could simply use the ratio of the
intercepts of the coordinate axes, but as the models exhibit a slight
peanut shape this gives a false impression of the shape of the
contours). We can see that the model is triaxial at the centre with
central values $(b/a)_{\rho}(r=r_0/10)\approx0.7$ and $(c/a)_{\rho}(r=r_0/10)\approx0.4$. For
$r>10r_0$, the potential is much more spherical, whilst the density is
still flattened. At these large radii, there is little mass. As the
total mass is finite, the behaviour at large radii mimics that of a
tracer population in a monopole potential. The density contours in
$(x,z)$ are not elliptically shaped, but take on a slight peanut
shape. \cite{Emsellem2011} shows that the slow rotators in the
ATLAS-3D sample (i.e. possible triaxial galaxies) have a boxiness
ratio $a_4/a$ consistent with pure elliptical or slightly boxy. Those
that have slightly discy contours appear to have a kinematically
decoupled core. It is therefore interesting that peanut-shaped
distributions arise in the models (and also in the models of
\cite{Schwarzschild1979}), but not in nature. However, we should note
that the degree of triaxiality presented here is larger than many
galaxies are believed to have. Also, we will see later that in
projection these models look elliptical.

\begin{figure}
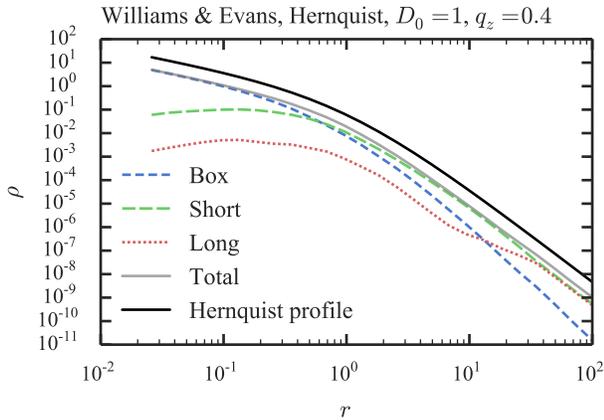

$$\includegraphics{{{figs/Fig3}}}$$
\caption{Contributions to the density in the triaxial WEH model along
  a radial ray with $\phi=\upifour$, $\theta=\upifour$. The boxes
  contribute most at the centre, whilst the loop orbits contribute
  equally in the outskirts where the potential is roughly
  spherical. Also shown is a true Hernquist profile offset by a factor of
  three for visibility. The positions are in units of $r_0$ and the
  density is in units of $M/r_0$.}
\label{SplitWE}
\end{figure}

In Figure~\ref{SplitWE}, we show the density profile along a radial
ray at the spherical polar angles $\phi=\upifour$ and
$\theta=\upifour$ decomposed in terms of the orbit classification.
We see that in the centre of the model, the majority of the density is contributed by the box orbits whilst in the outskirts the long and short axis loops contribute equally.
The orbits are classified based on their limits in the St\"ackel fudge
scheme as described in \cite{SandersBinney2015_Triax}. A small fraction of orbits have boundaries that do not correspond to any orbital class in a St\"ackel potential so we label them as approximately box, short-axis loop or long-axis loop depending on the relative magnitude of their actions. Our orbit classification method artificially imposes regularity on the orbital contribution to the model as some orbits will be chaotic but incorrectly given a regular label. There exist several methods for ascertaining whether an orbit is regular or chaotic (e.g. Lyapunov exponents -- see \cite{Vasiliev2013} for a nice summary or the small alignment index [SALI] \cite{Skokos2003} that was applied recently to characterise the orbits in an axisymmetric galactic potential by \cite{Zotos2014}) but these require long orbit integration. It seems for non-rotating galactic potentials of interest the majority of weight is on orbits that appear regular over a Hubble time (see Section~\ref{Discussion} for more discussion) such that our orbital decomposition is meaningful.

\begin{figure*}
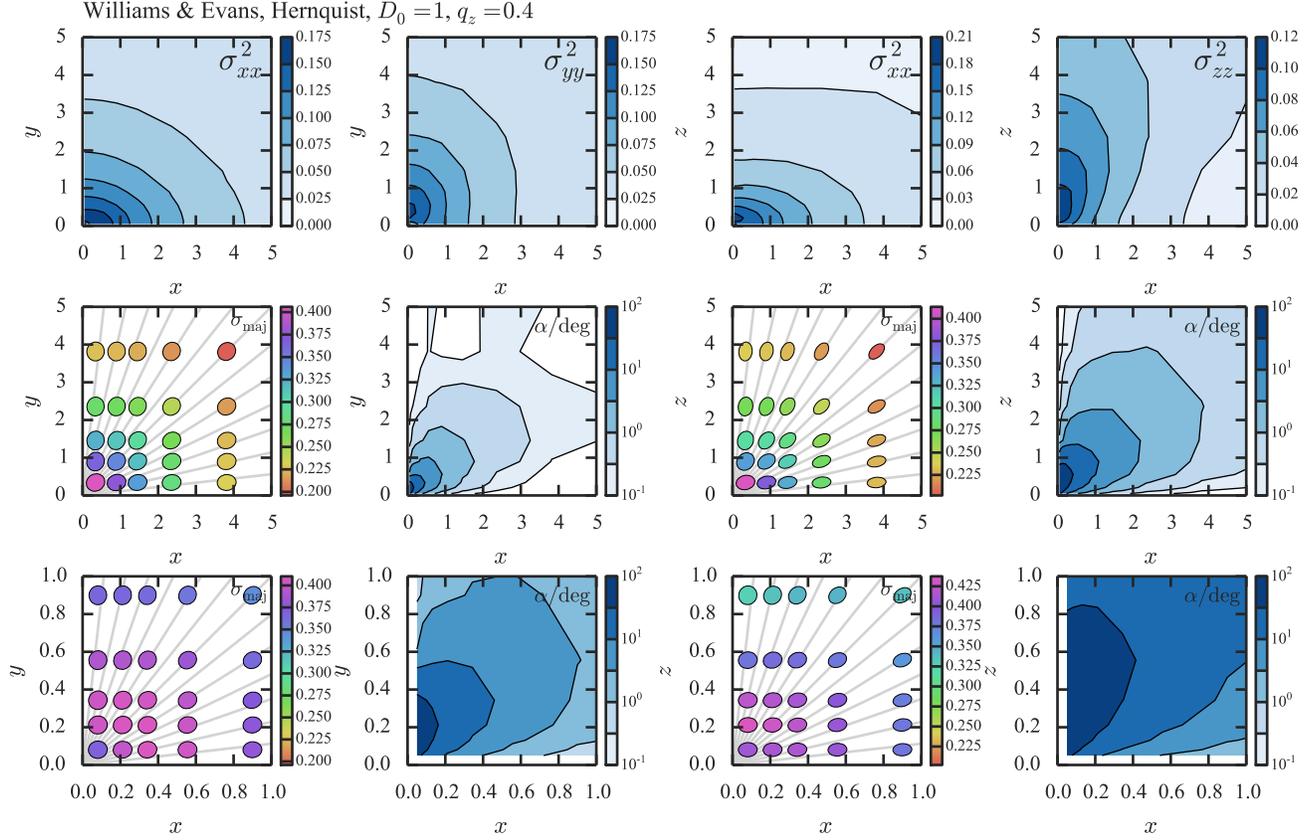

$$\includegraphics{{{figs/Fig4}}}$$
\caption{Velocity dispersions and ellipses of triaxial WEH model: in
  the top row the first panel shows $\sigma^2_{xx}$ in the $(x,y)$
  plane, the second panel shows $\sigma^2_{yy}$ in the $(x,y)$ plane,
  the third panel shows $\sigma^2_{xx}$ in the $(x,z)$ plane and the
  fourth panel shows $\sigma^2_{zz}$ in the $(x,z)$ plane. In the
  second row, the first panel shows the velocity ellipses in the
  $(x,y)$ plane, the second panel shows the tilt with respect to the
  radial direction (drawn in light grey in the first panel) in the
  $(x,y)$ plane, the third panel shows the velocity ellipses in the
  $(x,z)$ plane and the fourth panel shows shows the tilt with respect
  to the radial in the $(x,z)$ plane. The colours show the magnitude
  of the velocity dispersion along the major axis of each ellipse and
  only the shape of the ellipse is important. The sizes of the
  ellipses are arbitrary. The bottom row shows a zoom-in of the second row.
  The positions are in units of $r_0$ and the dispersions are in units
  of $\sqrt{GM/r_0}$.}
\label{VelDispWE}
\end{figure*}

In Figure~\ref{VelDispWE}, we show the velocity dispersions in the
$(x,y)$ and $(x,z)$ planes. $\sigma^2_{xx}$ and $\sigma^2_{yy}$
produce approximately elliptical contours whilst $\sigma^2_{zz}$
produces a narrow waisted distribution in the $(x,z)$ plane with the
dispersion falling off much more slowly along the $z$ axis. We also
plot the velocity ellipses in the same two planes. Clearly the
velocity ellipses are very radial for $z>0$. In the $(x,z)$ plane, the
ellipses are near radially-aligned for most of the explored space. At
large radii, the ellipses become more radially aligned and at small
radii they appear to become aligned with the Cartesian coordinates. In
the $(x,y)$ plane, we have a similar situation but the ellipses are
near circular so any alignment of the ellipses in this plane is
subtle. This configuration is seen in models with St\"ackel potentials as the velocity ellipsoids are aligned with the confocal ellipsoidal coordinate system. Fig.~\ref{SplitWE} shows that at the centre of the model the box orbits dominate so the coordinate system is naturally aligned with Cartesian coordinates whilst beyond the scale radius loop orbits become more significant and the velocity ellipses naturally become more spherical aligned. Although we have used a St\"ackel-based approach to estimate the actions, we have allowed the focal distances of the coordinate system to be a function of energy so we believe that this configuration is not a consequence of our calculation but a genuine feature of the models

\subsection{Binney's flattened isochrones}

The spherical isochrone density-potential pair is given by
\begin{equation}
\begin{split}
\rho_{\rm I}(r) &= \frac{M}{4\upi}\frac{3(r_0+r_a)r_a^2-r^2(r_0+3r_a)}{r_a^3(r_0+r_a)^3},\\
\Phi_{\rm I}(r) &= -\frac{GM}{r_0+r_a},
\end{split}
\end{equation}
where $r_a=\sqrt{r_0^2+r^2}$ \citep{BinneyTremaine}. The isochrone density profile has a
constant core and falls off as $r^{-4}$ at large radii. The
isochrone potential is convenient as it is the most general potential
in which the actions can be analytically calculated \citep{Ev90}. The
Hamiltonian as a function of the actions is
\begin{equation}
H_{\rm I}(\bs{J}) = -\frac{(GM)^2}{2[J_r+\half(L+\sqrt{L^2+4GMr_0})]^2}.
\end{equation}

\cite{Henon1960} demonstrated via Eddington inversion that the
isotropic distribution function is given by
\begin{equation}
\begin{split}
f_{\rm I}(&H_{\rm I}) = \frac{\mathcal{N}}{\sqrt{2}(2\upi)^3(GMr_0)^{3/2}}\frac{\sqrt{\mathcal{H}}}{[2(1-\mathcal{H})]^4}\Big[27-66\mathcal{H}+320\mathcal{H}^2
\\&-240\mathcal{H}^3+64\mathcal{H}^4+3(16\mathcal{H}^2+28\mathcal{H}-9)\frac{\sin^{-1}\sqrt{\mathcal{H}}}{\sqrt{\mathcal{H}(1-\mathcal{H})}}\Big],
\end{split}
\end{equation}
where
\begin{equation}
\mathcal{H} = -\frac{H_{\rm I}r_0}{GM}.
\end{equation}
Note here $\mathcal{N}=1$ but we retain it as it is necessary when adjusting the \df.
\cite{Binney2014_ISO} proposed a scheme to flatten these models such that
axisymmetric analogues of the isochrone could be constructed. The
density of the model in planes of constant energy is adjusted by
scaling each of the actions by $\alpha_i$. The new model is given by
\begin{equation}
f_{\rm B}(\bs{J}) = \alpha_r\alpha_\phi\alpha_\theta f_{\rm I}(\alpha_rJ_r,\alpha_\phi J_\phi,\alpha_\theta J_\theta).
\end{equation}
In order to retain the spherically-averaged density profile of the
isochrone, the $\alpha_i$ are not independent. We are free to choose
two constants $\alpha_{\phi0}$ and $\alpha_{\theta0}$. The
coefficients are given by
\begin{equation}
\begin{split}
\alpha_r(\bs{\bar J})&=(1-\psi)\alpha_0+\psi\alpha_{r0},\\
\alpha_\phi(\bs{\bar J})&=(1-\psi)\alpha_0+\psi\alpha_{\phi 0},\\
\alpha_\theta(\bs{\bar J})&=\alpha_{\theta 0},
\end{split}
\label{alphas}
\end{equation}
where
\begin{equation}
\begin{split}
\alpha_0(\bs{\bar J})&=1-\frac{\Omega_L}{\Omega_L+\Omega_r}(\alpha_{\theta0}-1),\\
\alpha_{r0}(\bs{\bar J})&=1-\frac{\Omega_L}{\Omega_r}(\alpha_{\phi0}+\alpha_{\theta0}-2),\\
\psi(\bs{\bar J})&=\tanh(\bs{\bar J}/\sqrt{GMr_0}).
\end{split}
\end{equation}
In these expressions, the frequencies, $\Omega_i$, are evaluated in the
spherical isochrone potential at the actions $\bs{\bar J}=(\bar J,\bar
J,\bar J)$ which is the barycentre of a plane of constant energy given
by\footnote{Note there is a typo in equation (11) of \cite{Binney2014_ISO} such that the
  expression inside the square root should read
  $\frac{(GM)^2}{-2E}+3GMr_0$, not $\frac{(GM)^2}{-2E}-3GMr_0$.}
\begin{equation}
\bar{J} = \third\Big(\frac{2GM}{\sqrt{-2E}}-\sqrt{\frac{(GM)^2}{-2E}+3GMr_0}\Big).
\end{equation}
In this equation $E$ is evaluated using $H(\bs{J})$ for the spherical isochrone.
Expressions for the frequencies are given in
\cite{BinneyTremaine}. The linear combination in
equation~\eqref{alphas} was introduced by \cite{Binney2014_ISO} to produce
models that did not become prolate at the centre. In this way
$\alpha_r$ tends to $\alpha_\phi$ at low barycentric action.

\subsubsection{Example model}\label{IsoModel}

We now construct a triaxial generalization of one of these models. In
order to induce triaxiality, we require $\alpha_{\theta0}>1$ such that
the model is flattened in the $z$ direction and
$\alpha_{r0}<\alpha_{\phi0}$ such that the model is radially
biased. However, we also require $\alpha_{\phi0}+\alpha_{\theta0}<3$
such that $\alpha_{r0}>0$ everywhere.

\begin{figure*}
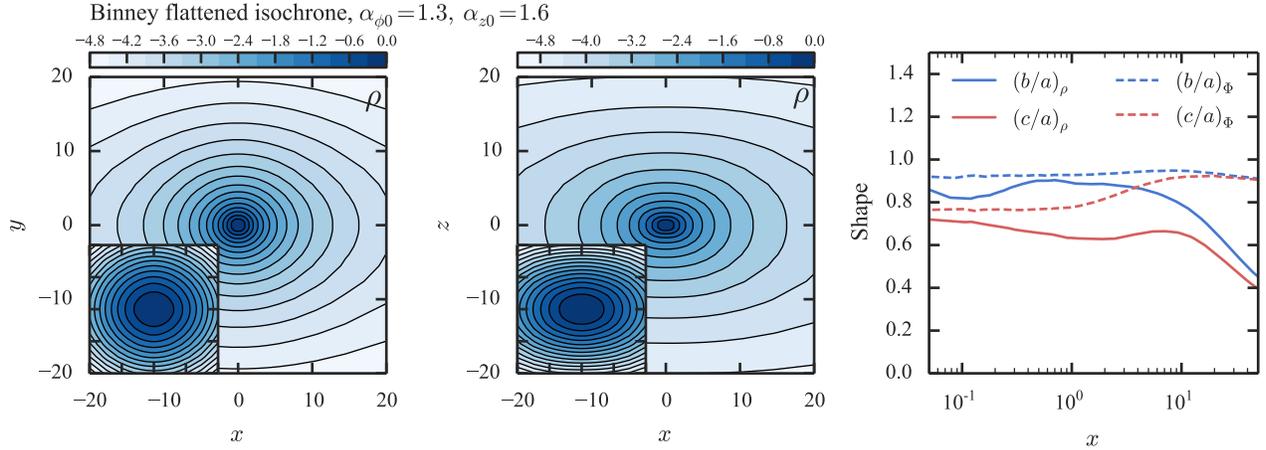

$$\includegraphics{{{figs/Fig5}}}$$
\caption{Density and shape of triaxial Binney isochrone model: the
  left panel shows a slice of the density in the $(x,y)$ plane and the
  middle panel shows the equivalent in the $(x,z)$ plane. The blue
  colours indicate the base $10$ logarithm of the density relative to
  the central value. The insets show zoom-ins for $|x|<1,y<|1|$ and
  $|x|<1,|z|<1$. In the right-hand panel we show the axis ratios of the
  fitted ellipses in these two planes with solid lines [$b/a$
  corresponding to $(x,y)$ and $c/a$ $(x,z)$]. The dashed lines show
  the equivalent for the potential. The positions are in units of
  $r_0$ and the density in units of $M/r_0^3$.}
\label{DensityB}
\end{figure*}

We initially tried the $\alpha_{\phi0}=1, \alpha_{\theta0}=1.4$ model
presented in \cite{Binney2014_ISO} but this produced a near-axisymmetric
model. We instead opted for $\alpha_{\theta0}=1.3$ and
$\alpha_{\phi0}=1.6$ and we set $G=M=r_0=1$. This produced a weakly triaxial model. In Figure~\ref{DensityB}, we plot the density contours of the model in the $(x,y)$ plane and
$(x,z)$ plane along with the axis ratios of the best-fitting ellipses
to the density and potential contours. We can see that model is
triaxial at the centre with $(b/a)_{\rho}(r=r_0/10)\approx0.8$ and
$(c/a)_{\rho}(r=r_0/10)\approx0.7$. For $r>10r_0$ the potential is tending
towards spherical whilst the density is still flattened. We see the
strong core in the density contours of the model and the slight boxy
shape of the density contours in the $(x,z)$ plane.


\begin{figure}
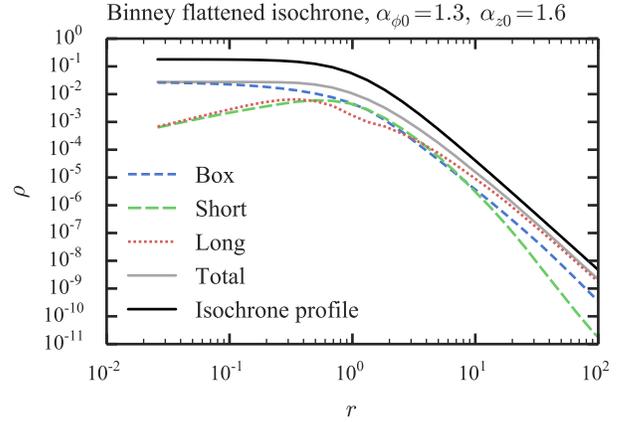

$$\includegraphics{{{figs/Fig6}}}$$
\caption{Contributions to the density in the triaxial Binney (2014)
  isochrone model along a radial line with $\phi=\upifour$,
  $\theta=\upifour$. Also shown is an isochrone profile offset by a
  factor of three for visibility. The positions are in units of $r_0$
  and the density is in units of $M/r_0$.}
\label{SplitB}
\end{figure}

In Figure~\ref{SplitB}, we show the density profile along a radial
line at the spherical polar angles $\phi=\upifour$ and
$\theta=\upifour$ decomposed in terms of the orbit classification. At
the centre, nearly all the density is contributed by the box
orbits. The contribution to the density from the loop orbits turns
over around $r=r_0$ due the harmonic core of the potential. For $r_0<r<3r_0$ the short-axis loop orbits and the box orbits are contributing equally whilst beyond $r=3r_0$ the long-axis loop orbits dominate. In the outskirts, the
long-axis loops contribute most with very weak contributions from the box orbits and the short-axis loops. This combination is
clearly necessary to produce the observed triaxial shape in the
outskirts. It is interesting to note the similarities and differences from the
Hernquist model. In both cases the density of box orbits outweights the loop orbits by two orders of magnitude. However, for the isochrone model the long and short axis loops contribute equally whilst for the Hernquist model the long axis loop orbit contributes over an order of magnitude less density than the short axis loop orbits. At large radii the box orbits contribute weakly in both cases with the loop orbits contributing most to the density. However, in the Hernquist model the short and long axis loop orbits are contributing equally at large radii whilst in the isochrone model the long-axis loop orbits are dominating the density. Note that in this case, unlike with the WEH model, the total density profile differs from an isochrone profile. The models will only exactly match for small $|\alpha_i-1|$ which is not satisfied for this model as $\alpha_R\approx3-\alpha_\phi-\alpha_z\approx0.1$. However, the inner and outer slopes match well.

In Figure~\ref{VelDispB}, we show the velocity dispersions in the
$(x,y)$ and $(x,z)$ planes. $\sigma^2_{xx}$ and $\sigma^2_{yy}$
produce approximately elliptical contours within $r\approx 2r_0$ as with
the Hernquist case. However, the velocity dispersions at the centre do
not rise nearly as steeply as the in the Hernquist
case. Interestingly, the $\sigma^2_{xx}$ contours break from
ellipticity for $y>3r_0$ as the outermost contour shown has a positive
curvature. Also $\sigma^2_{yy}$ seems to have a narrow waisted
distribution for $x\ga3r_0$. As in the Hernquist case, $\sigma^2_{zz}$
produces a narrow-waisted distribution in the $(x,z)$ plane with the
dispersion falling off much more slowly along the $z$ axis. We also
plot the velocity ellipses in the same two planes. Clearly the
velocity ellipses are very radial in both planes. The structure of the
radial alignment in both planes is similar to the Hernquist case with
the largest deviation from radial alignment occurring at $x=0$ and
small $y$ and $z$.

\begin{figure*}
$$\includegraphics{{{figs/Fig7}}}$$
\caption{Velocity dispersions and ellipses of triaxial Binney
  isochrone model: in the top row, the first panel shows
  $\sigma^2_{xx}$ in the $(x,y)$ plane, the second panel shows
  $\sigma^2_{yy}$ in the $(x,y)$ plane, the third panel shows
  $\sigma^2_{xx}$ in the $(x,z)$ plane and the fourth panel shows
  $\sigma^2_{zz}$ in the $(x,z)$ plane. In the second row, the first
  panel shows the velocity ellipses in the $(x,y)$ plane, the second
  panel shows the tilt with respect to the radial direction (drawn in
  light grey in the first panel) in the $(x,y)$ plane, the third panel
  shows the velocity ellipses in the $(x,z)$ plane and the fourth
  panel shows shows the tilt with respect to the radial in the $(x,z)$
  plane. The colours show the magnitude of the velocity dispersion
  along the major axis of each ellipse and only the shape of the
  ellipse is important. The sizes of the ellipses are arbitrary. The bottom row shows a zoom-in of the second row. The
  positions are in units of $r_0$ and the dispersions are in units of
  $\sqrt{GM/r_0}$.}
\label{VelDispB}
\end{figure*}

\section{Applications of models}\label{Applications}

We present two applications of our new triaxial models.

\subsection{Isophote twisting}

One of the main lines of evidence for the triaxiality of some
elliptical galaxies is the observation of isophote twisting
\citep{Be81,Emsellem2007}. Isophote twisting is the change in the
position angle of the major axis of the isophotal contours. In fast
rotating galaxies or bulges~\cite[e.g.,][]{Stark1977}, this could be attributed to an inner bar/disc
structure, whereas in slow rotators, the phenomenon is often
attributed to triaxiality~\citep[e.g.][]{Wi79}.  The projection of
a triaxial model with ellipsoidal axis ratios varying with distance
from the centre naturally produces gradual isophotal twists. Another
explanation for the isophote twisting is an intrinsic twisting
possibly due to a recent interaction.  Several authors have studied
the statistics of isophote twisting of large samples of galaxies
\citep{Carter1978,BenacchioGalletta1980,Leach1981}. \cite{FasanoBonoli1989}
analysed a sample of galaxies and concluded that tidal interactions
only played a small role in isophote twisting but that many of the
galaxies studied had evidence of a central spheroid.

Irrespective of the exact origin of isophote twisting, the range of
models able to reproduce this phenomenon is poor.  The occurrence of
isophotal twists was one of the main motivations for development of
triaxial St\"ackel models. So, it was a surprise when \cite{Franx1988}
showed that these models -- unusually for triaxial systems -- do not
produce isophote twisting at any viewing angle. Therefore, we have
hitherto been limited to constructing fully numerical (Schwarzschild
or M2M) models to explore this phenomenon.

\begin{figure}
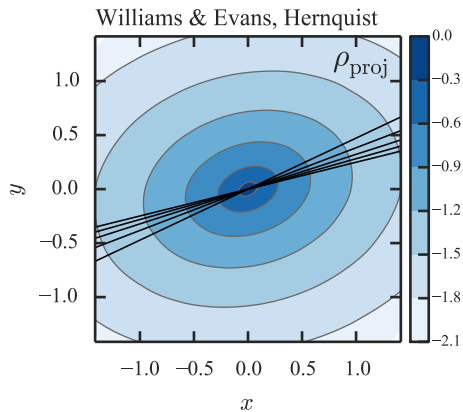

$$\includegraphics{{{figs/Fig8}}}$$
\caption{Logarithmic projected density of triaxial WEH model: the
  model is viewed along a unit vector with spherical polar angles
  $(\phi,\theta)=(\upifour,\upifour)$. The lines show the major axis
  of ellipses fitted to the four innermost contours to indicate the
  clockwise twisting of the density contours as we move outwards. The
  positions are in units of $r_0$ and the unit of density is
  $M/r_0^3$.}
\label{IsophoteTwist}
\end{figure}

Here, we demonstrate that the model explored in Section~\ref{WEModel}
exhibits isophote twisting. In Figure~\ref{IsophoteTwist}, we show the
projected density for the triaxial WEH model analysed in
Section~\ref{WEModel} when observed along a unit vector at spherical
polar angles $(\phi,\theta)=(\upifour,\upifour)$. We also show the
major axes of ellipses fitted to the four innermost density
contours. Clearly, the major axis of the density contours twists
clockwise as we move out from the centre. The gradient of the twist is
$\approx 8\degrees/r_0$. Recalling that the effective radius of a
Hernquist model is $\approx 1.8 r_0$, then the isophotal twisting
gradient is $\approx 14\degrees$/ effective radius, which is very
comparable to the data~\citep[see e.g., Figure 8 of][]{Leach1981}

\subsection{Radially unstable models}\label{RadialOrbitInstab}

The radial orbit instability has received much attention in the
literature from both theoretical and numerical studies (see
\cite{Merritt1999} for a nice review). The underlying physics is that
when spherical models contain a high fraction of radial orbits, the
models are unstable and tend to collapse to triaxial
distributions. \cite{PolyachenkoShukhman2015} present a discussion of the mechanism by which the instability operates. Some systems are unstable to fast-growing even and odd tangential Jeans instabilities due to the lack of tangential kinetic energy. Other systems, which have more centrally concentrated radial orbits, are unstable to slow-growing bar-like modes of the type described by \cite{LyndenBell1979} that act to align orbits. However, determining exactly when the instability will
set in has been difficult to assess.  Numerical experiments require
the construction of $N$-body realizations so are limited to specific
analytic equilibria such as Osipkov-Merritt $f(Q)$ models
\citep{MerrittAguilar1985,DejongheMerritt1988,MezaZamorano1997,Perez1996}
or polytropes $f\propto (-E)^pL^q$ \citep{Barnes1986}, or use models constructed via Schwarzschild's method. Recently, \cite{Antonini2009} and \cite{Vasiliev2012} have showed that triaxial models constructed via the Schwarzschild method are also susceptible to a radial-orbit instability when the central anisotropy of a Dehnen model with an inner density slope of $1$ rises to $\beta\approx0.4$. Theoretical studies have inspected the growing modes using the matrix method of
\cite{Kalnajs_Matrix}. \cite{PalmerPapaloizou1987} showed that all
radially-anisotropic polytropes were unstable, whilst
\cite{Saha1991,Saha1992} and \cite{Weinberg1991} derived conditions under which
$f(Q)$ models were radially unstable. The conclusions from both
approaches do not present a uniform picture and it seems difficult to
predict exactly when a model is unstable without significant
computation. Our new approach, however, provides another more general
way to construct models that could be used to shed more light on the
instability.

The degree to which the radial orbit instability is important in the
landscape of galactic formation and evolution today is unclear. Some
galaxies are believed to be weakly triaxial but the accretion of
baryons and formation of black holes may decrease the degree of
triaxiality at the centre of galaxies. Therefore, it is conceivable
that the radial orbit instability may be mainly of theoretical
interest and is often damped away in realistic systems. However, in
the hierarchical picture of galaxy formation a galaxy grows through
the accretion of smaller satellites which can be on very eccentric
radial orbits. Therefore, a typical anisotropy profile is isotropic at
the centre tending towards radial in the outskirts. It is therefore
conceivable that the radial orbit instability sets in during the
evolution and so is one of the mechanisms controlling the shape of
galaxies.

\begin{figure}
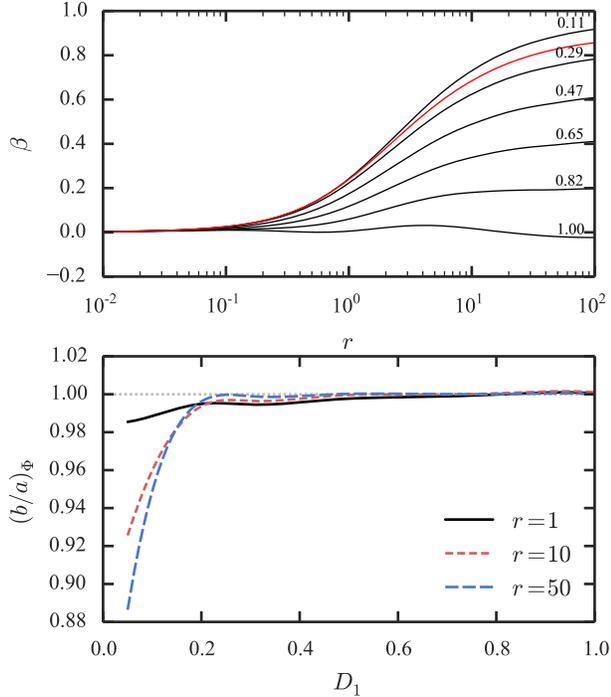

$$\includegraphics{{{figs/Fig9}}}$$
\caption{Radially unstable models: the top panel shows anisotropy
  profiles for seven spherical WEH models labelled by the parameter, $D_1$, that controls the outer anisotropy. Spherical radius $r$ is in units of
  $r_0$. $D_1=1$ is the model presented in Williams \& Evans (2015)
  and the other models use identical parameters apart from $D_1$.  The unlabelled red line corresponds to $D_1=0.2$, which is the critical model that separates the weakly spherical and triaxial endpoints. The
  bottom panel shows the shape of the potential contours in the
  $(x,y)$ plane at three different major axis $r$ values against $D_1$
  for models allowed to relax to triaxiality. }
\label{BetaProfiles}
\end{figure}

\begin{figure*}
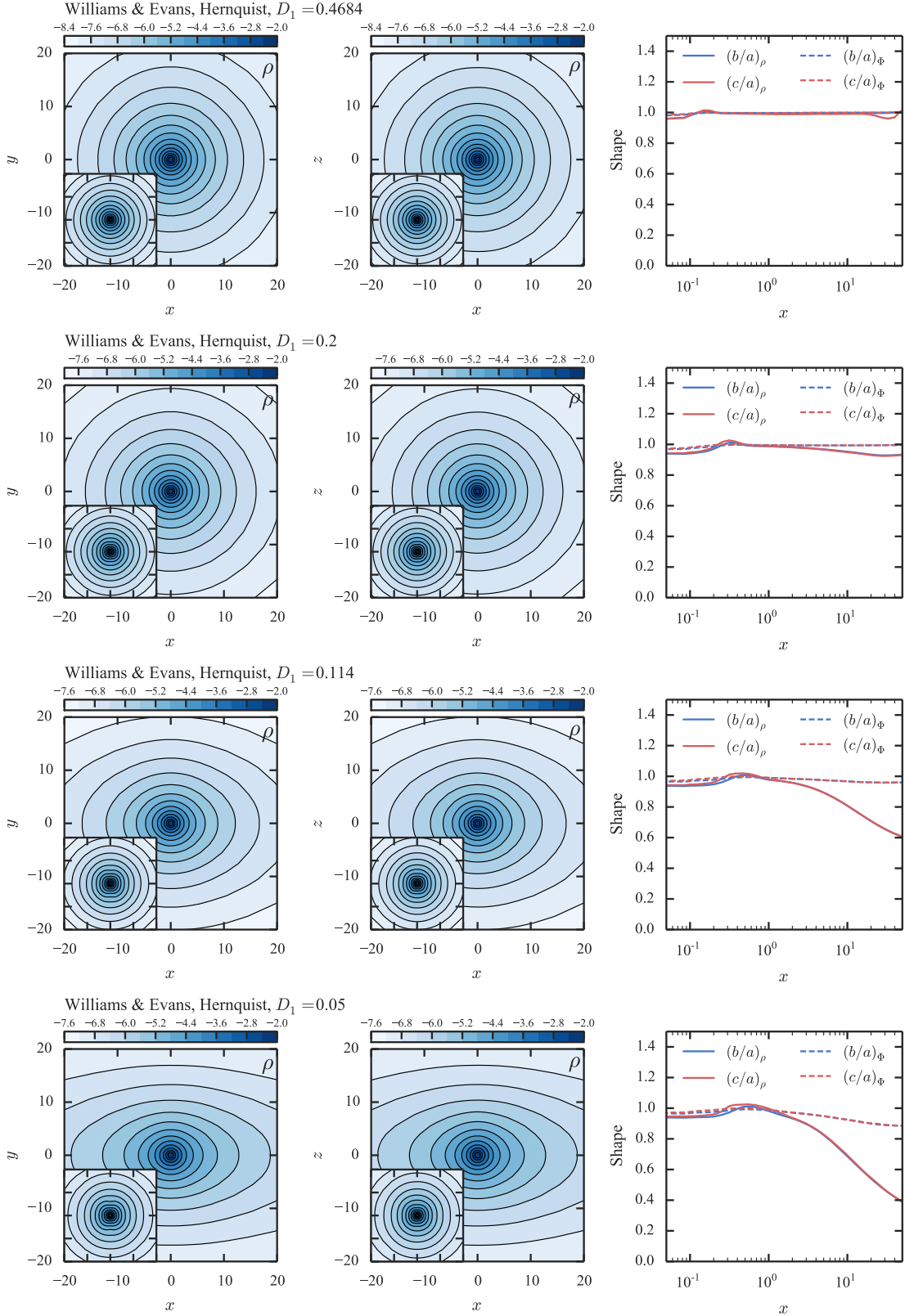

$$\includegraphics[width=0.85\textwidth]{{{figs/Fig10a}}}$$
\vspace{-0.9cm}
$$\includegraphics[width=0.85\textwidth]{{{figs/Fig10b}}}$$
\vspace{-0.9cm}
$$\includegraphics[width=0.85\textwidth]{{{figs/Fig10c}}}$$
\vspace{-0.9cm}
$$\includegraphics[width=0.85\textwidth]{{{figs/Fig10d}}}$$
\caption{Range of triaxial WEH models with variable outer
  anisotropy. Each row shows the results for models of differing $D_1$ which is noted above each row. The left two panels in each row show density slices in the $(x,y)$ and $(x,z)$ planes with the insets showing zoom-ins of $|x|<1,|y|<1$ and $|x|<1,|z|<1$. The right panels show the axis ratios of ellipses fitted to the density and potential contours in these two planes [$b/a$ corresponds to the $(x,y)$ plane and $c/a$ the $(x,z)$ plane]. For $D_1\ga0.2$ the models are spherical, whilst for
  $D_1\la0.2$ the models collapse to triaxial distributions. The
  positions are in units of $r_0$ and the density in units of
  $M/r_0^3$.}
\label{RangeofModelsD1}
\end{figure*}

We limit ourselves to models where $|J_\phi|$ and $|J_\theta|$ are on
an equal footing ($q_z=1$ in equation~\eqref{flattening}). These should produce spherical models as they only
depend on $J_r$ and $L$. However, we allow a transition to triaxiality
where the actions take on a slightly different meaning such that the
models are not necessarily spherical. We construct WEH models with
varying outer anisotropy by adjusting the parameter $D_1$. All other parameters are as
given in \cite{WilliamsEvans2015} isotropic Hernquist model, but $T_1$
is adjusted according to equation~\eqref{T0T1alteration}. In
Figure~\ref{BetaProfiles}, we show the anisotropy profiles for seven
models restricting the models to spherical symmetry. We show the
density contours and density and potential shapes in
Figure~\ref{RangeofModelsD1} of four of these models that have been given the freedom to relax to triaxiality. We see that the
$D_1=0.4684$ model is spherical. $D_1=0.2$ has a triaxial structure in
the outskirts of the density profile but the potential contours remain
approximately spherical. For $D_1=0.114$, the model has developed a
strong prolate shape in the outskirts and the potential contours are
prolate throughout the explored volume. When $D_1$ is reduced further
to $0.05$, the contours become more strongly prolate. In the lower
panel of Figure~\ref{BetaProfiles}, we show the $(b/a)_\Phi$ shape at
$r=r_0$, $r=10r_0$ and $r=50r_0$ against $D_1$. We see that beyond $D_1=0.2$ the models become prolate in nature. The anisotropy profile of the spherical equivalent of the critical $D_1=0.2$ model is shown in red in Figure~\ref{BetaProfiles}.

It appears that when $D_1\ga0.2$ the models remain approximately
spherical, whereas for $D_1\la0.2$ the models begin relaxing to a
prolate shape. This threshold corresponds to an anisotropy at the scale radius of $\beta(r_0)\approx0.25$. We interpret this result as follows: for $D_1\la0.2$
the spherical models are unstable and susceptible to the radial orbit
instability, such that the models relax to a radially-distended
density distribution. As we have adiabatically relaxed the models, we
are allowing instabilities that grow adiabatically slowly to
persist. However, we have only allowed the models to transition
through a range of triaxial models such that, although the models are
radially unstable, it is not clear that they will relax to the given
distributions. It may be that other equilibria can be explored on the
way to the final triaxial distribution.

In our approach the final models are approached via a series of
triaxial non-self-consistent equilibria. However, in reality when
$D_1\la0.2$ the models will pass quickly through a series of
non-equilibrium states to settle down to a final, possibly quite
different, triaxial endpoint. Therefore, we conclude that
$D_1\approx0.2$ is the transition point between stable and unstable
spherical models, and that for $D_1\approx0.2$, we can find
equilibrium models that are triaxial.

\cite{Antonini2009} and \cite{Vasiliev2012} found an approximate threshold for stability of triaxial Dehnen models constructed with the Schwarzschild method was $2T_r^2/T_t^2\approx1.4$ where $T_r$ is the radial kinetic energy and $T_t$ the tangential kinetic energy. Assuming a constant velocity anisotropy this threshold corresponds to $\beta\approx0.3$ (approximately the critical anisotropy at the scale radius) so seems consistent with the result found here.

Our method involves the adiabatic relaxation of the models. \cite{PolyachenkoShukhman2015} have shown that some models are susceptible to fast-growing radial modes that grow on timescales of order the radial period of the responsible orbits and \cite{Antonini2009} present models that have modes that grow on times of $\sim 20$ crossing times. As our models are generated adiabatically, fast-growing modes may be suppressed.

\section{Discussion}\label{Discussion}

The triaxial St\"ackel models are exactly integrable and so possess no
chaotic orbits. Our triaxial models are not integrable and will always
contain some chaotic orbits, which are nonetheless assigned
actions. In practice, the action label is meaningless and will vary
with time.  However, we have assigned the weight of this orbit based
on the \df\ $f(\bs{J})$, so the weight should correspondingly change in
time. However, it cannot, and so the model is not in equilibrium if it
contains chaotic orbits.

Chaotic orbits appear regular on short times but slowly diffuse
(Arnold diffusion). Arnold diffusion seems to be relatively slow
compared to a Hubble time such that the equilibrium will be valid on
scales of physical interest. However, at the centre of the models many
of the orbits can diffuse very rapidly from one semi-regular orbit to
the next such that the equilibrium is only valid on very short
timescales~\cite[e.g.,][]{Va98}. Therefore, for our approach to hold
good, we require the impact of chaotic orbits to be minimal otherwise
the model will not be in true equilibrium.

In fact, all equilibrium construction approaches (including
Schwarzschild or M2M) struggle to handle chaotic orbits. In the case of
Schwarzschild modelling, one is limited to integrating all orbits for
a fixed period of time such that a chaotic orbit can never fully be
incorporated. Chaotic orbits have a single integral of motion, the energy, such that a model of solely the chaotic orbits can be constructed as $f_c(E)$. \cite{Me96} adopt the approach of summing the densities of a series of chaotic orbits integrated for $100$ dynamical times at each energy to construct this component. A full Schwarzschild model is then the combination of this chaotic piece plus an additional regular piece that can be included in the usual way. We could choose to follow the same procedure here by segregating chaotic orbits and constructing $f(\bs{J})$ models for the regular region of the phase-space. In this respect, our modelling approach does not seem inferior to
other approaches and is probably equivalent.

Several authors have investigated the impact of chaotic orbits on numerically constructed Schwarzschild models. \cite{Voglis2002} and \cite{CapuzzoDolcetta} showed that, although non-rotating triaxial models contained $20-30\percent$ of chaotic orbits by mass, only a few percent of these orbits by mass exhibited Arnold diffusion over a Hubble time. The impact of the chaotic orbits is a slow evolution of the axis ratios of the halos over time as corroborated by \cite{ZorziMuzzios2012} and \cite{Vasiliev2013}. However in contrast to this, \cite{PoonMerritt2002} have constructed models of triaxial nuclei with central black holes with considerable chaotic contributions which have fixed axis ratios over several dynamical times. When figure rotation is included the fraction of chaotic orbits by mass can increase significantly to $\sim50\percent$ as shown by \cite{Voglis2006} and \cite{ManosAthanassoula2011}. In conclusion although many orbits are formally chaotic in realistic non-rotating galactic potentials only a small fraction are chaotic on timescales of physical interest and these act to only weakly alter the shapes of the resulting models. Therefore, we conclude that the impact of chaos on the results presented in this paper is weak.

\section{Conclusion}\label{Conclusions}

The range of triaxial equilibria for stellar systems remains largely
uninvestigated. The variety of intrinsic shapes and anisotropies for
which stable dynamical equilibria exist is not known, and the role of
figure rotation and tumbling is completely unexplored. One reason for
this is the paucity of tools for building such models.

Much of our insight into triaxial equilibria comes from the St\"ackel
or separable potentials made famous by \cite{deZeeuw1985a}. Here,
there are four orbital families, and self-consistent models can be
built by orbit superposition or Schwarzschild's method
\citep{Statler1987}. There is strong evidence that the presence of four
orbital families gives ample opportunity for exchanging orbits of
different shapes, whilst keeping the total density unchanged
~\citep{Hu92}. Therefore, the range of possible triaxial St\"ackel
equilibria may be very rich. Other than St\"ackel models, there are
only a handful of idealised triaxial equilibria in the
literature~\citep{Fr66,Va80}. Of course, Schwarzschild (1979) and
made-to measure~\citep{SyerTremaine1996} techniques do in principle
allow for a systematic exploration of triaxial equilibria. In
practice, these techniques have been so far largely limited to
axisymmetric systems because a smaller investment in computer time is
required to sample phase-space in this case.

Here, we have developed a new way of studying triaxial systems. Many
of the tools to do this exist in the literature, including algorithms
for the swift calculation of
actions~\citep{Binney2012,SandersBinney2015_Triax} and possible
ansatzes for forms of the distribution functions (\df s) in terms of
the actions for both cored~\citep{Binney2014_ISO} and
cusped~\citep{WilliamsEvans2015,Posti2015} galaxies. Synthesizing these
results, we have shown how to build triaxial models with analytic \df
s for the isochrone and Hernquist density laws. Their velocity
ellipsoids are aligned with Cartesian coordinates near the centre, but spherical polar coordinates at large radii. In terms of orbital families, box
orbits make a reasonable contribution at all radii in isochrone models,
but they are less important at large radii in Hernquist models.

We have studied two particular features of the models. First, they can
exhibit isophote twisting. This occurs when the position angle of the
major axis of the isophotes changes with radius, with changes of the
order of $10^\circ$ over an effective radius being
typical~\citep{Leach1981}. By projecting a triaxial Hernquist model
built with a \df\ using the Williams \& Evans (2015) ansatz, we
demonstrated that our models could reproduce such a phenomenon. This
is particualrly valuable, as the St\"ackel models -- unusually for
triaxial systems -- do not exhibit isophote
twisting~\citep{Franx1988}.  Secondly, we used our models to explore
the radial orbit instability. We considered the adiabatic relaxation of a
family of \df s with central isotropy and variable outer anisotropy governed by a parameter $D_1$. The endpoints of the
relaxation may change from spherical to triaxial for some critical
value of $D_1$. Although the evolution is constrained, and so the
endpoints may not be realised in practice, this nonetheless does give
us the critical value at which the radial orbit instability sets in.
We have given a particular example of this procedure for the triaxial
Hernquist models.

There are some further directions and possible applications to
pursue. The study of triaxial equilibria could be advanced by the
combination of this work with $N$-body simulations or possibly
made-to-measure models. As suggested in \cite{Binney2014_ISO}, these
models can be used to create seeds for $N$-body simulations. These
could then be evolved to investigate stability or the effects of
growth of a disc or black hole in the model. It has been shown by \cite{Antonini2009} that equilibrium models constructed with Schwarzschild models can be dynamically unstable and it is expected that some models constructed via the method presented here will also be unstable. The only way to address this issue is through $N$-body modelling. Additionally, triaxiality
seems an important effect to account for in the interpretation of
observations on external galaxies. To date only the Schwarzschild
models of \cite{vandenBosch2008} have successfully modelled a triaxial
elliptical galaxy. Therefore, one immediate application is to
construct fits to ATLAS-3D data for slow rotators which may exhibit
signatures of triaxiality.  Finally, our models can be generalized to
include streaming motion. The integrals of motion depend upon $v_i^2$
such that the sign of $v_i$ is irrelevant. We are therefore able to
give every particle a positive velocity such that the model has a net
streaming velocity. Note that this would affect both the kinematical
properties and the stability of the models constructed.

\section*{Acknowledgements}
JLS acknowledges the support of the Science and Technology Facilities
Council and thanks Angus Williams for useful conversations.

\bibliography{bibliography}
\bibliographystyle{mn2e}

\appendix
\section{Choice of coordinate system}\label{Coords}
In this appendix we detail how we choose the coordinate parameters
$\alpha$ and $\beta$ when using the St\"ackel fudge algorithm of \cite{SandersBinney2015_Triax}.
Given a general triaxial potential the only integral of motion we are in a position to calculate is the
energy, $E$. Therefore, we choose to make the choice of $\alpha$ and
$\beta$ a function of $E$ and we estimate the parameters using the closed loop orbits at each energy.

We construct a logarithmically spaced grid in energy from $E_0=\Phi(0,y_{\rm min},0)$ to $E_1=\Phi(0,y_{\rm max},0)$ with
$N_E = 24$ grid points. We choose $y_{\rm min}=0.015r_0$ and $y_{\rm max}=180r_0$ where $r_0$ is the scale of the target model. To find the short-axis closed loop orbit at
each energy we launch particles at a location $y_i$ along the $y$-axis
with velocity $v_{xi}=\sqrt{2E-2\Phi(0,y_i,0)}$ in the $x$-direction. We then record the $y$ value
of the next time the orbit crosses the $y$-axis ($y=y_f$) and the
corresponding $x$-velocity $v_{xf}$. We repeat this procedure and
minimise
\begin{equation}
\Omega_E^2(-y_i-y_f)^2+(-v_{xi}-v_{xf})^2,
\end{equation}
where $\Omega_E$ is the frequency of a circular orbit with energy $E$
in a spherical potential with the radial profile corresponding to the
profile along the $x$-axis in the triaxial potential. The same
procedure is employed to find the long-axis loop orbits by replacing
$x$ with $z$ in the above equations.

The long and short axis closed loop orbits are confined to the $(y,z)$
and $(x,y)$ planes respectively. In a St\"ackel potential they follow
elliptical lines of constant $\lambda$ given by
\begin{equation}
\begin{split}
\frac{y^2}{\lambda+\beta}+
\frac{z^2}{\lambda+\gamma} &= 1,\\
\frac{x^2}{\lambda+\alpha}+
\frac{y^2}{\lambda+\beta} &= 1.
\end{split}
\end{equation}
With the closed orbits in a general potential found we can find the
axis intercepts for the short axis loop: $(x_{\rm s}, y_{\rm s})$ and
the long-axis loop $(y_{\rm l},z_{\rm l})$. Fitting ellipses to these
points we find
\begin{equation}
\begin{split}
\beta &= \gamma - y_{\rm l}^2 + z_{\rm l}^2,\\
\alpha &= \beta - x_{\rm s}^2 + y_{\rm s}^2.
\end{split}
\end{equation}

In \cite{SandersBinney2015_Triax} we performed this procedure by
finding the closed loop orbits around the long and short axis and then
minimising the variation in the action describing the circulation of
the axis as a function of the coordinate parameters. This procedure
was unnecessary as, with the closed orbits, found the
coordinate parameters can be simply calculated without finding the
actions. The procedure presented here gives identical results to that
employed in \cite{SandersBinney2015_Triax} but is less cumbersome.

At some energies closed loop orbits cannot be found so we skip these
points and fill them using a linear extrapolation. Additionally, we
constrain $\alpha<\beta<\gamma$ such that $x$ is the long axis and $z$
the short axis. If our procedure produces estimates that violate this
(e.g. in a near-spherical potential) we set $\alpha=\beta-r_0^2/10$
and $\beta=\gamma-r_0^2/20$, where $r_0$ is the scale-length of our
models. Decreasing this arbitrary choice of $\gamma-\alpha$ and
$\gamma-\beta$ does not significantly affect the accuracy of the
calculations for radii $r<r_0$ but produces errors in the outskirts of
the models ($r\gg r_0$) as the focal length is significantly smaller
than the typical orbital scale. Our choice was selected such that the
spherical isochrone and Hernquist models were reproduced.

\section{Multipole expansion}\label{MultiExp}
To find the potential from the density of an $f(\bs{J})$ model we use a multipole expansion. Here we give details of this approach.
We evaluate the density on a 3D grid in spherical polar coordinates
$(r,\phi,\cos\theta)$. The grid in radius $r$ is logarithmically
spaced between two radii $r_{\rm min}$ and $r_{\rm max}$ such that
\begin{equation}
r_i = a_0\mathrm{e}^{\delta_i}+r_{\rm min},
\end{equation}
where
\begin{equation}
\delta_i = \frac{n_r}{N_r-1}\log(r_{\rm max}/r_{\rm min}),\quad n_r = 0,1,\dots,N_r-1.
\end{equation}
The scale $a_0$ is chosen to be the scale radius of the model of
interest, $r_0$, and we set $r_{\rm min}=0.01r_0$ and $r_{\rm max}=200r_0$ which encloses $~99\percent$ of the total mass in both the spherical Hernquist and isochrone models. The $\cos\theta$ and $\phi$ grids are based on $N_a$-point
Gauss-Legendre quadrature on the intervals $(-1,0)$ and $(0,\upi/2)$
respectively. We set $N_r=40$ and $N_a=15$.

The potential $\Phi$ is expanded as
\begin{equation}
\Phi(r,\phi,\theta) = -4\upi G\sum_{l=0,2,\dots}^{l=l_{\rm max}}\sum_{m=0,2,\dots}^{m=l}\phi_{lm}(r)Y_l^m(\phi,\theta),
\end{equation}
where $Y_l^m(\phi,\theta)$ are spherical harmonics. The sum is over
even values of $l$ and $m$ using a maximum $l$ of $l_{\rm max}=8$.
We choose to work with a symmetric definition of $Y_l^m(\phi,\theta)$
\citep{VasilievSMILE} given by
\begin{equation}
Y_l^m(\phi,\theta) = \frac{\mathcal{Y}}{\sqrt{2l+1}}\bar P_l^m(\cos\theta)\cos m\phi,
\end{equation}
where $\mathcal{Y}=\sqrt{2}$ if $m>0$ otherwise $\mathcal{Y}=1$. The
normalized associated Legendre polynomials, $\bar P_l^m(\cos\theta)$,
are calculated using the GNU Science Library \citep[GSL,][]{GSL} as
\begin{equation}
\bar P_l^m(\cos\theta) = \sqrt{\frac{2l+1}{4\upi}}\sqrt{\frac{(l-m)!}{(l+m)!}} P_l^m(\cos\theta),
\end{equation}
where $P_l^m(\cos\theta)$ are the associated Legendre polynomials. The
$Y_l^m(\phi,\theta)$ satisfy the orthogonality condition
\begin{equation}
\int_0^\upi\d\theta\int_0^{2\upi}\d\phi\,\sin\theta Y_l^mY_{l'}^{m'} = \frac{\delta_{mm'}\delta_{ll'}}{2l+1}.
\end{equation}

$\phi_{lm}(r)$ is found by integrating the potential of the series of
spherical shells of thickness $\delta r$ with mass $\rho_{lm}(r)\delta r$
inside and outside the radius $r$. Mathematically we have
\begin{equation}
\phi_{lm}(r) = r^{-l-1}I_{lm}^{(0)}(r)+r^lI_{lm}^{(\infty)}(r),
\end{equation}
where
\begin{equation}
\begin{split}
I_{lm}^{(0)}(r) &= \int_0^r\mathrm{d}a\,a^{l+2}\rho_{lm}(a),\\
I_{lm}^{(\infty)}(r) &= \int_r^\infty\mathrm{d}a\,a^{-l+1}\rho_{lm}(a),
\end{split}
\end{equation}
and
\begin{equation}
\rho_{lm}(r) = 8(-1)^{m/2}\int_{-1}^{0}\mathrm{d}(\cos\theta)\int_0^{\upi/2}\mathrm{d}\phi\,Y_l^m(\phi,\theta)\rho(r,\phi,\theta).
\end{equation}
The integral for $\phi_{lm}(r)$ is calculated using the trapezoidal
rule in the coordinate $\delta=\log((r_i-r_{\rm min})/a_0)$ and the
double integral for $\rho_{lm}(r)$ is found using Gauss-Legendre
quadrature.

Similarly, to find the forces we must evaluate
\begin{equation}
\frac{\upartial\phi_{lm}}{\upartial r} = -\frac{l+1}{r^{l+2}}I_{lm}^{(0)}(r)+lr^{l-1}I_{lm}^{(\infty)}(r),
\end{equation}
and the derivatives with respect to $\phi$ and $\theta$ are found by
analytically differentiating $Y_l^m(\phi,\theta)$.

We begin the calculation of these quantities by evaluating the density
on the grid in $(r,\phi,\theta)$. This is the costliest part of the
calculation but it can be parallelized. Then for each $l$ and $m$ we
evaluate $\rho_{lm}(r)$ on the grid in $r$ and use these results to
compute $\phi_{lm}(r)$ and $\upartial\phi_{lm}/\upartial r$ on a grid
in $r,l$ and $m$. Assuming a finite mass model (i.e. $\rho_{lm}(r)$
has a shallower divergence than $r^{-3}$ as $r\to0$ and falls off
steeper than $r^{-3}$ as $r\to\infty$) the integral from $r=0$ to
$r=r_{\rm min}$ is found as $I_{lm}^{(0)}(r_{\rm min}) =
\half\rho_{lm}(r_{\rm min})r_{\rm min}$. We set $I_{lm}^{(\infty)}=0$
for $r>r_{\rm max}$.

For $r<r_{\rm min}$ we use quadratic extrapolation of $\phi_{lm}(r)$
and linear extrapolation of $\upartial\phi_{lm}/\upartial r$. For
$r>r_{\rm max}$ we assume the density is zero and so extrapolate
$\phi_{lm}(r)$ and $\upartial\phi_{lm}/\upartial r$ as
\begin{equation}
\begin{split}
\phi_{lm}(r>r_{\rm max}) &= \phi_{lm}(r_{\rm max})\Big(\frac{r_{\rm max}}{r}\Big)^{l+1},\\
\frac{\upartial \phi_{lm}}{\upartial r}(r>r_{\rm max}) &= \phi_{lm}(r_{\rm max})\Big(\frac{r_{\rm max}}{r}\Big)^{l+2}.
\end{split}
\end{equation}

The potential and forces calculation was tested using a triaxial
perfect ellipsoid St\"ackel potential \citep{deZeeuw1985a} as well as
a flattened Hernquist model (replacing the square of the spherical
radius $r^2$ with $m^2=x^2+(y/q_y)^2+(z/q_z)^2$ in the Hernquist
density profile) for which the potential and forces were calculated
using equation (2.140) of \cite{BinneyTremaine} for the potential of a
general density distribution stratified on concentric ellipsoids.

\section{Cross-check with generating function approach}\label{CrossCheck}
In the main section of the paper we have used the triaxial St\"ackel fudge approach of \cite{SandersBinney2015_Triax} to construct the models due to its speed. However, the speed comes at the cost of increased errors in the actions. Using the St\"ackel fudge the action estimates will oscillate. For some orbits this oscillation is genuine (see Section~\ref{Discussion} on resonant and chaotic orbits), whilst for others it is a natural consequence of the approximation used. This naturally will introduce errors in the calculations to construct the models. \cite{SandersBinney2015_Triax} explored this and discussed the error the St\"ackel fudge method produced when finding the moments of similar non-self-consistent models. It was found that the Jeans equation was satisfied to $\la10\percent$ for two tracer \df s in a triaxial NFW potential.

\cite{SandersBinney2014} presented an alternate method for estimating actions in triaxial potentials. This algorithm will converge on the true action when one exists (for regular orbits), but is unfortunately much slower as it relies on orbit integration. However it can be used here as a cross-checking tool. Here we summarise the method and present a calculation of the density of one of our models using this approach.

\subsection{Generating function construction}
The generating function method of \cite{SandersBinney2014} is based on the construction of orbital tori methods pioneered by \cite{McGillBinney} and operates by construction of a generating function from the angle-actions in a simple analytic potential to those in the target potential from a series of orbit samples. Here we very briefly detail the method.

Given a target potential and a initial $(\bs{x},\bs{v})$ coordinate we integrate the orbit. Here we choose to integrate for $T = 8t_{\rm orb}$ where $t_{\rm orb}$ is the orbital period of a circular orbit with the same orbital energy in a spherical potential with a radial profile corresponding to the $x$-axis profile of the triaxial potential. From this orbit we sample $N_T = 200$ points and infer the orbital type (loop or box) from the angular momentum. We then choose a corresponding toy potential (triaxial harmonic oscillator for the box orbits and isochrone for the loops). We estimate the parameters of the toy potentials by fitting the toy forces to the true forces at the maximum $(|x|,|y|,|z|)$ points for the harmonic oscillator and by fitting the toy forces to the true forces at pericentre and apocentre for the isochrone potential (note this differs from the procedure employed in \cite{SandersBinney2014} who minimised the variance of the toy Hamiltonian for the orbit samples). With the toy potential estimated we find the toy actions and angles $(\bs{J}',\btheta')$. The true actions are then related to these toy variables via the generating function $S(\btheta',\bs{J})$ as
\begin{equation}
\bs{J} = \bs{J}'-2\sum_{\bs{n}}\bs{n}S_{\bs{n}}\cos\bs{n}\cdot\btheta',
\end{equation}
where $S_{\bs{n}}$ are unknown Fourier components of the generating function. From our orbit samples we have $N_T$ such equations which can be solved for $(\bs{J},S_{\bs{n}})$ by minimising the sum of the square differences between the toy actions and those calculated using the generating function series. We consider terms up to $|\bs{n}|\leq N_{\rm max}=6$. In addition to the checks of the solutions employed in \cite{SandersBinney2014} to ensure good coverage of all modes, we also ensure the radial action for the loop orbits is positive, all actions for the box orbits are positive and the average of the action over the corresponding angle does not deviate from the estimated action by more than the maximum estimated action. If these criteria are not satisfied we increase the number of samples (if there is less than one sample per $\pi$ phase of $\bs{n}\cdot\btheta'$ for any mode $\bs{n}$) or the time integration window (if $\bs{n}\cdot\btheta'$ does not wrap at least $2\pi$ for any mode $n$) by a factor of two until an upper limit (four times the initial choice) and if the conditions are still not satisfied we return the average actions.


\subsection{A cross-check}
\begin{figure}
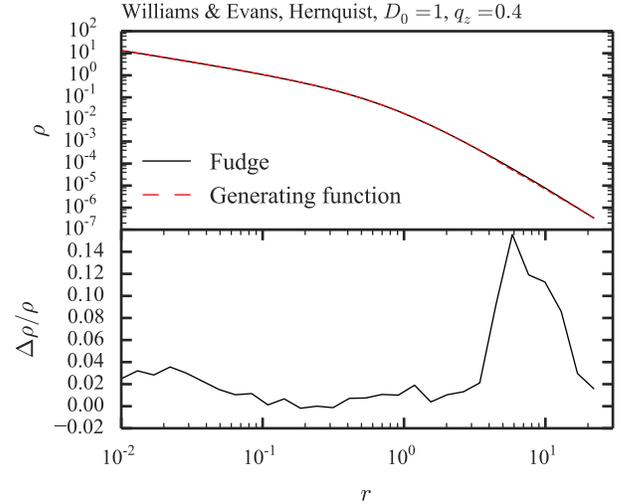

$$\includegraphics{{{figs/FigA1}}}$$
\caption{Density profile of triaxial WEH model along a ray with spherical polar angles $(\phi,\theta)=(\upifour,\upifour)$ calculated using
the triaxial St\"ackel fudge method (black solid) and the generating function method (red dashed). The lower panel shows the relative difference in the density. The positions are in units of $r_0$ and the unit of density is $M/r_0^3$.}
\label{dens_comp}
\end{figure}

Here we perform a cross-checking calculation for two of our models to measure the error the St\"ackel fudge approach introduces. In Figure~\ref{dens_comp} we plot the density profile of the converged WEH model of Section~\ref{WEModel} along a ray at spherical polar angle ($\phi=\upifour,\theta=\upifour$) calculated using the fudge method and the generating function method, and in Figure~\ref{dens_comp_iso} we show the same calculation for the isochrone model of Section~\ref{IsoModel}. Note the full convergence process was not performed using both approaches. We simply took the converged potential from the St\"ackel fudge calculation and re-calculated the density in this potential using the generating function approach. For the WEH model the relative error in the density is $\lesssim10\percent$ everywhere. The largest errors occur outside $r\approx5r_0$. For the isochrone model the relative error in the density is $\lesssim2\percent$ for $r<5r_0$ but the error steeply rises beyond this up to $30\percent$ for $r=20r_0$. It is unclear whether this systematic discrepancy is due to the fudge or generating function calculation but it appears to occur in the regions where the long-axis loop orbits are prominent so perhaps is due to their mishandling. We note that inspecting a less extremal model ($\alpha_\phi=1.3,\alpha_z=1.4$) we find that the two methods are agree well.

\begin{figure}
$$\includegraphics{{{figs/FigA2}}}$$
\caption{Density profile of triaxial isochrone model along a ray with spherical polar angles $(\phi,\theta)=(\upifour,\upifour)$ calculated using
the triaxial St\"ackel fudge method (black solid) and the generating function method (red dashed). The lower panel shows the relative difference in the density. The positions are in units of $r_0$ and the unit of density is $M/r_0^3$.}
\label{dens_comp_iso}
\end{figure}

\label{lastpage}
\end{document}